\documentclass[proof]{pasj00}
\begin{document}
\SetRunningHead{Author(s) in page-head}{Running Head}
%\Received{}%{yyyy/mm/dd}
%\Accepted{2011/05/07}%{yyyy/mm/dd}
%\Published{2011/10/25}%{yyyy/mm/dd}

\title{Widely Extended [OIII] 88 {\micron} Line Emission 
	around the 30 Doradus Region Revealed with AKARI FIS-FTS}

%%% begin:list of authors
% Do NOT capitalize all letters in "textsc".
\author{Mitsunobu \textsc{Kawada} 
  \thanks{Present Address: Institute of Space Astronautical Science (ISAS) 
	/ JAXA, 3-1-1 Yoshinodai, Chuo-ku, Sagamihara 252-5210, Japan; 
	e-mail: kawada@ir.isas.jaxa.jp},
%}
%\affil{Present Address: ISAS/JAXA 3-1-1 Yoshinodai, Chuo-ku, Sagamihara 
%	252-5210,  Japan}
%\email{kawada@ir.isas.jaxa.jp}
%\author{
  Ai \textsc{Takahashi},
  Akiko \textsc{Yasuda},
  Yuichi \textsc{Kiriyama},
  Tatsuya \textsc{Mori},
  Akio \textsc{Mouri},
  Hidehiro \textsc{Kaneda}}
\affil{Graduate School of Science, Nagoya University, Furo-cho, Chikusa-ku, 
	Nagoya 464-8602, Japan}
\author{Yoko \textsc{Okada}}
\affil{I. Physikalisches Institut, Universit\"at zu K\"oln, 
	Z\"ulpicher Str. 77, 50937 K\"oln, Germany}%\email{bbbbb@xxx.xxx.xx.xx}
\author{Hidenori \textsc{Takahashi}}
\affil{Gunma Astronomical Observatory, 6860-86 Nakayama, Takayama, 
	Agatsuma, Gunma 377-0702, Japan}%\email{bbbbb@xxx.xxx.xx.xx}
\and
\author{Noriko {\sc Murakami}}
\affil{Bisei Astronomical Observatory, 1723-70 Okura, Bisei-cho, Ibara-shi, 
	Okayama 714-1411, Japan}%\email{ccccc@xxx.xxx.xx.xx}
%%% end:list of authors

%%% Please use the following style in case that sorting by 
%%% affilation is impossible. 
%
% \author{%
%   D-Firstname \textsc{D-Familyname}\altaffilmark{1}
%   E-Firstname \textsc{E-Familyname}\altaffilmark{1,2}
%   and
%   F-Firstname \textsc{F-Familyname}\altaffilmark{2}}
% \altaffiltext{1}{Address of Institute}
% \email{ddddd@xxx.xxx.xx.xx}
% \email{eeeee@xxx.xxx.xx.xx}
% \altaffiltext{2}{Address of Institute}

%% `\KeyWords{}' always has to be placed before `\maketitle'.
 %Do NOT move this preamble from here!

\maketitle

\begin{abstract}
We present the distribution map of the far-infrared [OIII]~88{\micron} 
line emission around the 30 Doradus (30 Dor) region in the Large Magellanic 
Cloud obtained with the Fourier Transform Spectrometer of the Far-Infrared 
Surveyor onboard AKARI. The map reveals that the [OIII] emission is widely 
distributed by more than $10'$ around the super star cluster R136, implying 
that the 30 Dor region is affluent with interstellar radiation field hard 
enough to ionize O$^{2+}$. The observed [OIII] line intensities are as high 
as $(1-2)\times10^{-6}$ W m$^{-2}$ sr$^{-1}$ on the peripheral regions 
$4'-5'$ away from the center of 30 Dor, which requires gas densities of
$60-100$ cm$^{-3}$. However the observed size of the distribution of 
the [OIII] emission is too large to be explained by massive stars in the 30 
Dor region enshrouded by clouds with the constant gas density of 
$10^2$ cm$^{-3}$. Therefore the surrounding structure is likely to be 
highly clumpy.  We also find a global correlation between the [OIII] and 
the far-infrared continuum emission, suggesting that the gas and dust are 
well mixed in the highly-ionized region where the dust survives in clumpy 
dense clouds shielded from the energetic photons.
\end{abstract}

\section{Introduction}

Star forming regions and interstellar media have been well studied in all 
the wavelength regimes from radio to $\gamma$-ray.  An advantage using 
far-infrared (far-IR) light is that we can trace photo-ionization and 
dissociation regions without needing to correct for extinction.  In the far-IR 
region, there are important forbidden lines of atoms and ions, such as 
[OI]~63{\micron},~146{\micron}, [OIII]~88{\micron}, 
[NII]~122{\micron},~205{\micron}, and [CII]~158{\micron}, which are good 
tracers of the interstellar media in various gas phases. 

30 Doradus (30 Dor) in the Large Magellanic Cloud (LMC) is one of the most 
famous massive star forming regions (e.g. \cite{Kennicutt1984}). Its 
distance is about 50 kpc and the physical scale of $1'$ corresponds to 15 pc. 
In the center, as revealed by the Hubble Space Telescope, there are a lot of 
massive stars in a compact region, which is named the R136 super star cluster 
(\cite{Selman1999}; \cite{Massey1998}). The bubble of $1'$ in diameter 
exists around the center of R136, where gas is almost completely evacuated 
(\cite{Indebetouw2009}; \cite{Rubio1998}).  Prominent X-ray and H$\alpha$ 
filamentary and/or shell structures are extended in the north and the west 
direction from the center \citep{Townsley2006}. In the optical emission 
lines such as [SII] $\lambda$6724{\AA} and [OIII] $\lambda$5007{\AA}, 
widely extended components of ionized gas were revealed around the 30 Dor 
region by the University of Michigan/CTIO Magellanic Cloud Emission Line 
Survey (MCELS; \cite{Smith2000}). 

From the early era of IR astronomy, the 30 Dor region has been studied by 
various IR observations. Using the FIFI instrument of the Kuiper Airborne 
Observatory (KAO), \citet{Poglitsch1995} observed the far-IR forbidden lines 
of [CII]~158{\micron} and [OI]~63,~146{\micron} in the 30 Dor region, and 
made a line intensity map of [CII]~158{\micron} with an arc-minute spatial 
resolution over $7'\times 7'$ area. Balloon-borne IR telescopes, such as 
Balloon-borne Infrared Carbon Explore (BICE), observed the LMC region.  BICE 
mapped the entire area of the LMC by the [CII] line with a beam of about 
$15'$ size. \citet{Rubin2009} discussed the photoelectric heating and [CII] 
cooling in the LMC.  The first powerful tool for IR spectroscopy was provided 
by the Infrared Space Observatory (ISO) operated in the late 1990s. 
The ISOCAM made line intensity maps of [NeII]~12.81{\micron} and 
[NeIII]~15.56{\micron} emission lines for the central $\sim 3'$ area of 
the 30 Dor region \citep{Madden2006}.   The ionization potentials of Ne 
and Ne$^{+}$ are 21.56 and 40.96 eV, respectively.  Therefore, the [NeII] 
and [NeIII] lines are good tracers of highly-ionized gas, similarly to the 
[OIII] line with the ionization potential of 35.12 eV.  In the far-IR, the 
ISO/LWS observed more than 10 points with different physical conditions 
for the 30 Dor region (\cite{Vermeij2002a}; \cite{Vermeij2002b}). However, 
the spatial information provided by the ISO/LWS was rather poor, because 
these observations were performed by a single beam.

For technical reasons, far-IR spectral mapping observations with high 
spatial resolution were difficult. After ISO, AKARI takes a unique position 
in the far-IR spectroscopy.  The Japanese IR astronomical satellite, AKARI 
\citep{Murakami2007}, which was launched on February 21, 2006 (UT), has two 
kinds of focal-plane instruments; Infrared Camera (IRC) \citep{Onaka2007} 
and Far-Infrared Surveyor (FIS) \citep{Kawada2007}, to cover the whole IR 
regime from 2 to 180~{\micron}.  Far-IR spectroscopic capability of AKARI 
is realized by a Fourier transform spectrometer (FTS) as a function of the 
FIS \citep{Kawada2008}. The advantage of FIS-FTS over the ISO/LWS is an 
imaging spectroscopic capability with higher spatial resolution; 
each pixel of the FIS-FTS detector arrays can simultaneously take a spectrum 
with about 40$''$ (FWHM) angular resolution, while the ISO/LWS has the 
single beam of 80$''$ in FWHM.  The physical scales corresponding to their 
resolutions in the LMC are about 10 pc and 20 pc for FIS-FTS and the ISO/LWS, 
respectively.

Spectral maps with high spatial resolution are essential to investigate the 
interstellar media, especially those around ionized regions.  The structure 
of an ionized region is determined by the locations of ionizing stars and 
the distribution of the ambient interstellar media.  High spatial resolution 
maps of the far-IR lines clearly show the structure of star forming regions 
relating to the environment of star formation. Spitzer Space Telescope made 
spectral line maps of 30 Dor by using the IRS and the MIPS in a SED mode in 
the Spitzer legacy survey of the LMC: Surveying the Agents of a Galaxy's 
Evolution (SAGE; \cite{Meixner2006}), which revealed the ionization state 
of the 30 Dor region and super star cluster R136 \citep{Indebetouw2009}.  
AKARI FIS-FTS observed many local regions distributed around the 30 Dor 
region and provided their spectral maps.  In this paper, we present the 
large-scale spatial distribution of the [OIII]~88{\micron} line emission 
around the 30 Dor region. 

The advantage of the AKARI [OIII] data over the ISOCAM [NeIII] and the 
Spitzer data is the spatial coverage of much wider areas, i.e., those far 
outside the bubbles around 30 Dor.  Thus our data provide information on 
the ISM influenced by intense radiation from massive stars rather than 
shocks by stellar winds and supernovae in R136.  The optical 
[OIII] map with MCELS also covers a large area.  As compared to the far-IR 
[OIII] map, the optical map has a higher spatial resolution, but suffers 
extinction effects in dense gas regions, which may hamper accurate 
measurements of the change of the line intensity with the distance from 
R136.

\section{Observations}

AKARI carried out far-IR spectroscopic observations from April 2006 to 
August 2007. As a result, about 600 pointed observations were performed 
with FIS-FTS.  More than two thirds of the FIS-FTS observations were 
parallel observations with the IRC, in which the IRC has priority for 
observations.  Since the fields-of-view (FOVs) of the IRC and the FIS are 
separated by about half a degree, FIS-FTS observed off-primary target skies 
in the parallel observations.  Yet these data are useful as a spectroscopic 
survey of the diffuse interstellar medium, since many IRC observations 
concentrate on the Galactic plane and the LMC.

\begin{figure}
  \begin{center}
    \FigureFile(120mm,150mm){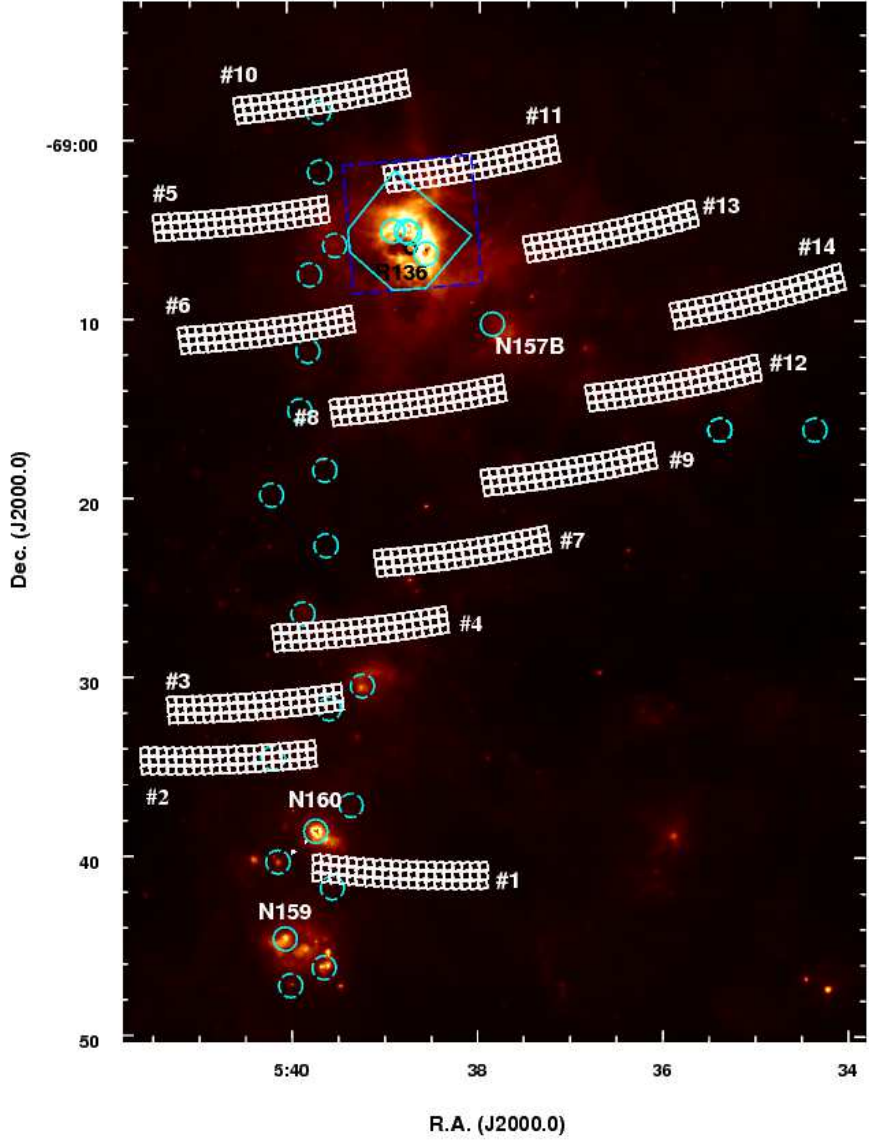}
    %%% \FigureFile(width,height){filename}
  \end{center}
  \caption{Observational areas with FIS-FTS overlaid on the Spitzer 
	MIPS~24{\micron} map from the SAGE archives. Each observing point 
	is indicated by the white frame in the $20\times3$ array format of 
	the short-wavelength-band detector labeled with the number listed 
	in table \ref{table:obs_log}.  
        The observational points of the ISO/LWS are also plotted by the open 
	circles with a size of the beam; 
	the line intensities from the solid circles are reported by 
	\citet{Vermeij2002a}, while those from the dashed circles are not 
	reported in any papers listed in the ISO archives database.
        The dashed square and the solid rectangle at the center of 30 Dor 
	indicate the coverage of spectral mapping with the Spitzer/IRC and 
	the MIPS in a SED mode, respectively \citep{Indebetouw2009}.}
  \label{fig:obs_area}
\end{figure}

In this paper, we selected 14 pointed observations around the 30 Dor regions, 
which targeted relatively bright regions
among the FIS-FTS observations of the LMC. An observation log is summarized 
in table \ref{table:obs_log}, and observed areas are shown in figure 
\ref{fig:obs_area}, plotted on the Spitzer/MIPS~24{\micron} map from the 
SAGE program.  For comparison, the observational positions of the ISO/LWS 
are marked by the small open circles with a size of the beam, and the area 
of the spectral mapping with the Spitzer/MIPS in a SED mode is indicated 
by the rectangle in the center of 30 Dor.

Although our observations do not cover the center of 30 Dor, they cover 
wide areas around the 30 Dor region and the south ridge.  As shown in 
figure \ref{fig:obs_area}, our observational areas are located on the 
periphery of the area covered with Spitzer and thus complementary to the 
Spitzer observations.  Each observing point has 60 pixels with a FOV of 
30$'' \times$ 30$''$ for each pixel in the short wavelength band, and 45 
pixels with a $50''$ $\times$ $50''$ pixel FOV in the long wavelength band.  
All the observational data analyzed in this paper are taken in a low 
resolution mode (named SED mode), whose spectral resolution is about 
1.2 cm$^{-1}$.

\begin{table}
  \caption{Observation log}\label{table:obs_log}
  \begin{center}
    \begin{tabular}{cccclc}
      \hline
      No. & date & position (J2000.0)\footnotemark[$*$] & obs. param.\footnotemark[$\dagger$] 
      	& comment\footnotemark[$\ddagger$] & det. limit\footnotemark[$\S$] \\
      \hline
      1 & 2006/10/14-21:15 & (84.74402, -69.67834) & SED, 0.5s & IRC04 (b;Ns) & 3.9 \\
      2 & 2006/10/19-21:57 & (85.20209, -69.57140) & SED, 0.5s & IRC04 (b;Ns) & 1.1 \\
      3 & 2006/10/21-21:54 & (85.13556, -69.52066) & SED, 0.5s & IRC02 (b;N) & 1.7 \\
      4 & 2006/10/23-18:32 & (84.85551, -69.45006) & SED, 0.5s & IRC02 (b;N) & 1.1 \\
      5 & 2006/10/23-20:11 & (85.16322, -69.07068) & SED, 0.5s & IRC04 (b;Ns) & 3.4 \\
      6 & 2006/10/24-21:00 & (85.09940, -69.17079) & SED, 0.5s & IRC02 (b;N) & 1.1 \\
      7 & 2006/10/25-18:29 & (84.58520, -69.37970) & SED, 0.5s & IRC02 (b;N) & 1.2 \\
      8 & 2006/10/25-21:48 & (84.70576, -69.23942) & SED, 0.5s & IRC02 (b;N) & 1.0 \\
      9 & 2006/10/26-20:57 & (84.30852, -69.30413) & SED, 0.5s & IRC02 (b;N) & 0.9 \\
      10 & 2006/10/26-22:36 & (84.95287, -68.95736) & SED, 0.5s & IRC02 (b;N) & 1.1 \\
      11 & 2006/10/27-20:06 & (84.56963, -69.02005) & SED, 0.5s & IRC02 (b;N) & 1.4 \\
      12 & 2006/10/27-21:45 & (84.03696, -69.22217) & SED, 0.5s & IRC02 (b;N) & 1.2 \\
      13 & 2006/10/29-20:03 & (84.20209, -69.08269) & SED, 0.5s & IRC02 (b;N) & 1.0 \\
      14 & 2006/10/30-20:51 & (83.81157, -69.14057) & SED, 0.5s & IRC02 (b;N) & 1.3 \\
      \hline
      \\
      \multicolumn{6}{@{}l@{}}{\hbox to 0pt{\parbox{140mm}{\footnotesize
        \footnotemark[$*$] Equatorial coordinates of the center of the short-wavelength-band array
        \par\noindent
        \footnotemark[$\dagger$]  Operational parameters of FIS-FTS; spectral 
        	resolution mode and reset interval (see \cite{Kawada2007})
        \par\noindent
        \footnotemark[$\ddagger$] Astronomical observation template of the IRC 
        	\citep{Onaka2007} in a parallel observation
        \par\noindent
        \footnotemark[$\S$] Typical 3$\sigma$ detection limit of the [OIII] 
	88{\micron} line [$\times 10^{-7}$ W m$^{-2}$ sr$^{-1}$]

      }\hss}}
    \end{tabular}
  \end{center}
\end{table}

\section{Data Reduction}

We used the standard reduction tools for FIS-FTS to derive spectra. Details 
of the standard tools are described in the AKARI FTS Toolkit 
Manual\footnote{AKARI (ASTRO-F) Observers Page 
-- http://www.ir.isas.jaxa.jp/ASTRO-F/Observation/}. 
Here, we briefly explain the way to obtain spectra from raw signal and to 
extract emission lines from the spectra.

Raw data are processed through the standard pipeline tools, in which a 
linearity correction, glitch removing and a correction of the optical 
path difference (OPD) are performed with an eye-ball check.  Then, a set of 
interferograms for each pixel is derived. Interferograms are transformed 
to raw spectra by applying a discrete Fourier transformation at the second 
stage of the pipeline, where the adopting OPD range and the apodizing 
function are selectable. In this paper, we applied a box-type apodizing 
function and used data in an OPD range of $\pm$~0.35~cm for the short 
wavelength band, a little shorter than the full range ($\pm$~0.42~cm), 
to avoid a channel fringe \citep{Kawada2008}.  The resulting spectral 
resolution of the short wavelength band is reduced to $\sim$~1.4~cm$^{-1}$ 
from $\sim$~1.2~cm$^{-1}$ of the full performance for the flat apodization.  
In the final step of the pipeline, raw spectra are converted to sky spectra 
by correcting for instrumental factors.  Details of the calibration scheme 
of FIS-FTS and the accuracy are described in \citet{Murakami2010}.

\begin{figure}
  \begin{center}
    \FigureFile(80mm,75mm){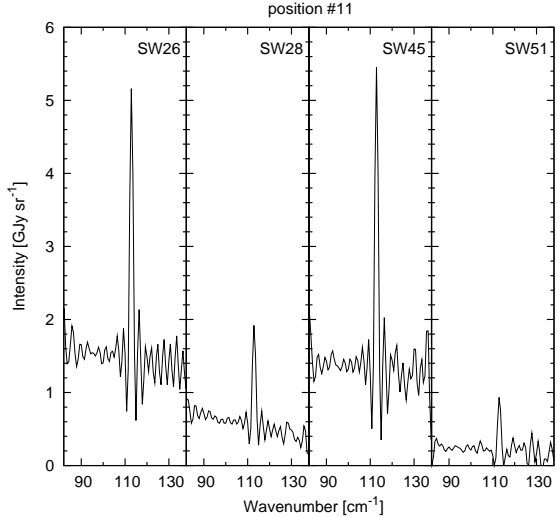}
    %%% \FigureFile(width,height){filename}
  \end{center}
  \caption{Examples of the spectra of 4 pixels in the short wavelength band 
  	obtained from the observation \#11.  The [OIII] line at 
  	113~cm$^{-1}$ (88~{\micron}) and continuum emissions with different 
  	slopes are clearly detected.}
  \label{fig:SW_spectrum}
\end{figure}

Since 30 forward and backward scans are obtained by a single pointed 
observation, there is a set of 60 spectra for each pixel.  From the 
set of the spectra, we calculated an average spectrum and estimated the 
error of each spectral element.  Figure \ref{fig:SW_spectrum} shows 
examples of the averaged spectra obtained in the short wavelength band, 
exhibiting the clear detection of the [OIII]~88{\micron} line at 
113~cm$^{-1}$ wavenumber.  The structure around the line indicates a sinc 
function due to the box-type apodizing function.  Dust continuum emissions 
are also detected; the SW26 and SW45 pixels show relatively flat spectra, 
while the continuum spectrum of the SW28 pixel indicates a cooler dust 
temperature.  From these spectra, we extracted [OIII] line intensities by 
simply subtracting continuum emissions from the signals integrated over 
111.0 -- 114.5~cm$^{-1}$.  The contribution of the continuum emissions at 
the line position is estimated from signals in both sides of the line, 
90.0 -- 111.0~cm$^{-1}$ and 114.5 -- 130.0~cm$^{-1}$.  The detection limit 
depends on the sensitivity of each pixel and also on the environmental 
condition of observations.  The 3$\sigma$ detection limits of the [OIII] 
line listed in table \ref{table:obs_log} are median values of array pixels 
for each pointed observation estimated from errors at the [OIII] line. The 
minimal 3$\sigma$ detection limit is about 
0.6~$\times$~10$^{-7}$~W~m$^{-2}$~sr$^{-1}$, and a typical value is about 
1~$\times$~10$^{-7}$~W~m$^{-2}$~sr$^{-1}$.

\begin{figure}
  \begin{center}
    \FigureFile(120mm,150mm){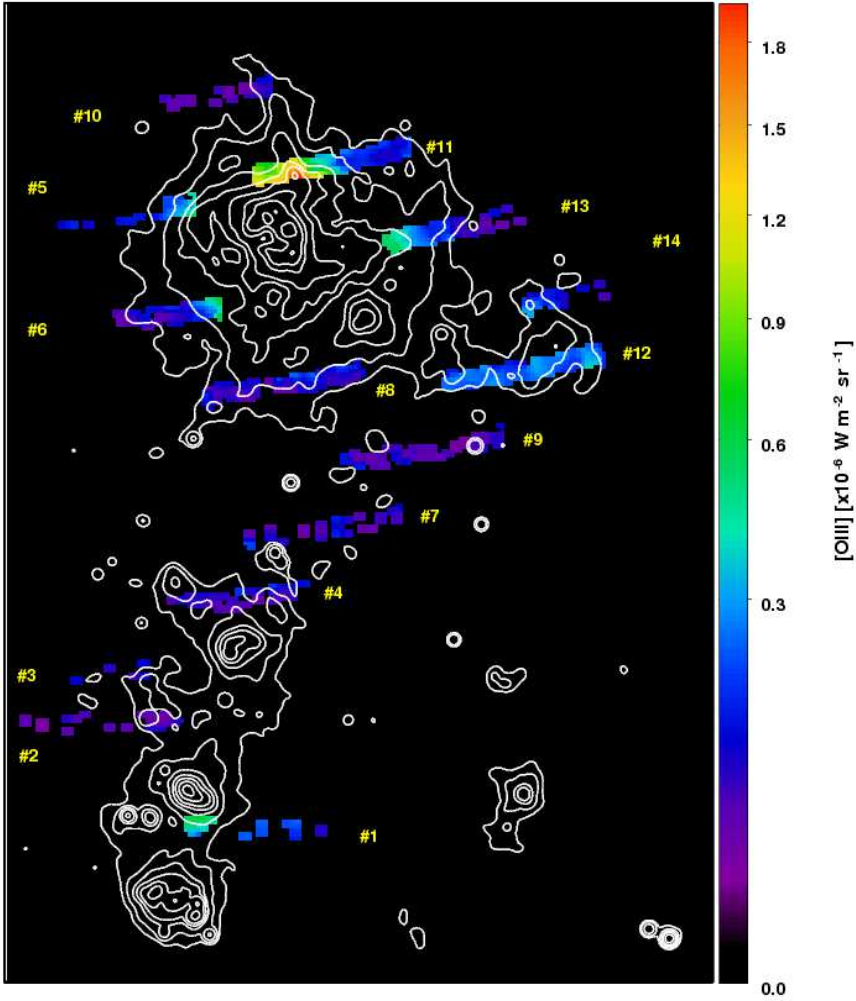}
    %%% \FigureFile(width,height){filename}
  \end{center}
  \caption{[OIII]~88{\micron} line intensity map, shown together with the 
  	MIPS~24{\micron} contour map whose spatial resolution is reduced 
  	to the same as the [OIII] map.
	}
  \label{fig:OIII_map}
\end{figure}

We also obtain spectra in the long wavelength band, which include the 
[CII]~158{\micron} line emission detected in many observations.  However, 
due to the non-uniformity of the detector performance among the pixels, 
the spectral data are still not well calibrated to discuss the spatial 
distribution of the [CII] line. Therefore, in this paper, we treat only the 
[OIII]~88{\micron} line, which traces highly-ionized gas. 
The [OIII]~88{\micron} intensity map thus obtained is shown in figure 
\ref{fig:OIII_map} together with the MIPS~24{\micron} intensity contour 
map. The [OIII] map is created by re-gridding pixel data on the 
astronomical coordinates and masking those with signal-to-noise ratios less 
than two.  The MIPS~24{\micron} contour map in the figure is convolved with 
a Gaussian kernel whose width match the spatial resolution of the FIS 
($\sim 40''$). The map clearly shows widely distributed [OIII]~88{\micron} 
line emission with arc-minute scale structures around 30 Dor.
From ISO archival data, we find that the ISO/LWS barely detects the 
[OIII]~88{\micron} line emission at the position in the FIS-FTS \# 3 area 
(see figure \ref{fig:obs_area}) with the intensity of about 2 $\times$ 
10$^{-8}$ W m$^{-2}$ sr$^{-1}$, which is  below our typical 3$\sigma$ 
detection limits 
as shown in table \ref{table:obs_log}.  The map in figure 
\ref{fig:OIII_map} is consistent with the ISO measurement.  This line 
intensity level represents that from quiescent parts of the LMC and the 
intensities of the spatially extended emission seen in figure 
\ref{fig:OIII_map} are $5-200$ times higher than this level.

\section{Results}

\subsection{[OIII] Intensity Map}

We significantly detect the [OIII]~88{\micron} emission at many pixels in 
each pointed observation.  Figure \ref{fig:OIII_map} shows the global 
structure of the [OIII] emission in about one square degree; the emission 
region of the [OIII] line is extended around the 30 Dor region more widely 
than the area covered with the Spitzer/MIPS in a SED mode.  The structure 
of the [OIII] line emission, the highly-ionized region, is well correlated 
with that of MIPS~24{\micron} (contours) tracing hot dust emission around 
the 30 Dor region.  In particular, a filamentary structure at the northern 
part seen in the MIPS~24{\micron} contours is also recognized well in the 
[OIII] intensity map (obs.\#10 \& \#11). Spitzer obtained 
spectra in the 30 Dor region with the IRS and the MIPS in a SED mode.  The 
[OIII]~88{\micron} intensity map taken by the MIPS covers the center of 30 
Dor (rectangle in figure \ref{fig:obs_area}), partially overlapped with our 
map at the north edge. 
The MIPS measurement shows values about three times larger than ours, but 
both are roughly consistent by considering large uncertainties of our absolute 
line intensity calibration \citep{Murakami2010}.  
Owing to our clear detection outside the Spitzer observing 
area, we reveal the widely extended structure of the highly-ionized region. 
In particular, the ionized regions are extended in the north and the west 
direction along the MIPS~24{\micron} structures, which have spatial 
correspondence to the X-ray and H$\alpha$ filamentary structures. 

On the other hand, the south-west region (obs.\#12 \& \#14) outside the 
30 Dor does not have a clear correlation between the [OIII]~88{\micron} and 
the MIPS~24{\micron} emissions. We also detect the [OIII] emission 
significantly from this region, whose distribution is rather flat as 
compared to the MIPS~24{\micron}.  This region contains the 30 Dor C 
superbubble, which is a large (45 pc radius) shell of material swept up by 
fast stellar winds and supernovae from tens of massive stars in an OB 
association.  There is a mixture of thermal and nonthermal X-ray 
emission (\cite{Bamba2004}; \cite{Townsley2006}, \cite{Yamaguchi2009}); 
any photoionization in this region may receive a contribution by hard 
radiation from the supernova remnants. There is probably also direct 
collisional ionization to O$^{2+}$.  
The south region far outside 30 Dor (obs.\#1 -- \#4, \#7 \& \#9) may have 
a different relation from the region near 30 Dor.
At least, the obs.\#1 data are apparently influenced by N159 and N160, other 
massive star forming regions (see figure \ref{fig:obs_area}).  
The MIPS~24{\micron} emission regions in the south are compact as 
compared to the 30 Dor region.  There are extended molecular clouds in this 
region and ionizing stars are located in dense cores, suggesting that star 
formation is in an early stage. 

For clarity of presentation, we categorize the observational points into 
four groups according to their positions with respect to 30 Dor. The NW 
group (obs.\#10, \#11 and \#13) is located in the north-west region of 30 
Dor and has a good correlation with the MIPS~24{\micron} surface brightness.
The SW group (obs.\#12 and \#14) is located on the far west side of 30 Dor 
and has a rather flat distribution of the [OIII] line emission.  The SE 
group (obs.\#5, \#6 and \#8) is located in the south-east region of 30 Dor.  
The SR group (obs.\#1 -- \#4, \#7 and \#9) is located along the south ridge 
of molecular clouds.

\subsection{Point to Point Correlations}

\begin{figure}
  \begin{center}
    \FigureFile(80mm,75mm){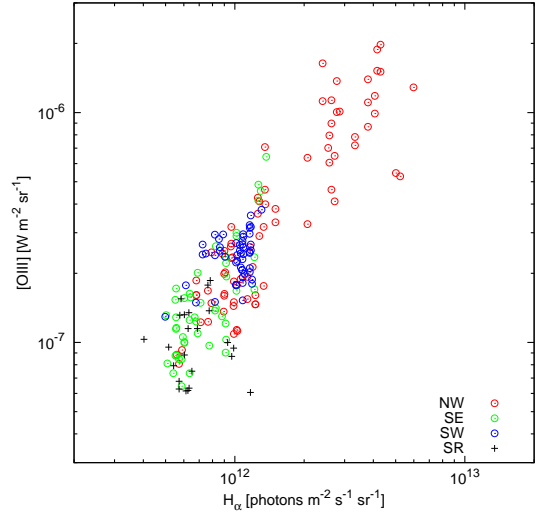}
    %%% \FigureFile(width,height){filename}
  \end{center}
  \caption{Scatter plot between the [OIII] line emission and the H$\alpha$ 
  	emission. The different marks and colors correspond to the data 
  	from the different regions of the observed area (see text). }
  \label{fig:corr_Ha}
\end{figure}

Figure \ref{fig:corr_Ha} shows a scatter plot between the 
[OIII]~88{\micron} line intensity and the H${\alpha}$ intensity, where the 
H$_{\alpha}$ data are taken from the Southern H-Alpha Sky Survey Atlas 
\citep{Gaustad2001}.  The different colors correspond to the data from the 
different group defined above. We can see a global correlation between the 
[OIII] and H${\alpha}$ emissions for a range of about one and a half 
orders of magnitude. The correlation, however, is not so tight at the higher 
line intensity region.  
A possible reason is difference in ionization degrees of 
the gases associated with the H${\alpha}$ and [OIII]~88{\micron} emission; 
much higher energies of photons ($>$ 35.12 eV) are required to ionize 
O$^{+}$ to O$^{2+}$ than to ionize H atom.

\begin{figure}
  \begin{center}
    \FigureFile(80mm,75mm){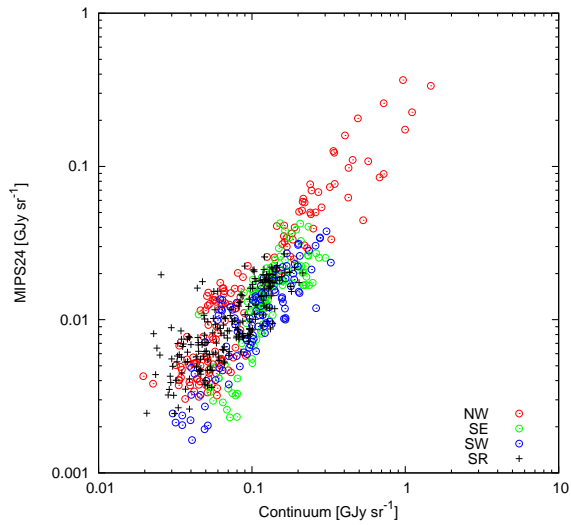}
    %%% \FigureFile(width,height){filename}
  \end{center}
  \caption{Scatter plot between the continuum emission at 88~{\micron} 
	detected with FIS-FTS and the MIPS~24{\micron} emission. The 
	spatial resolution of the MIPS~24{\micron} map is reduced to 
	matched the FIS-FTS map. The colors of the data points are the same 
	as in figure \ref{fig:corr_Ha}.}
  \label{fig:corr_mips}
\end{figure}

We also make a scatter plot between the continuum emission at 88~{\micron} 
derived with FIS-FTS and the MIPS~24{\micron} emission, as shown in figure 
\ref{fig:corr_mips}.  The spatial resolution of the MIPS~24{\micron} map is 
reduced to match the FIS-FTS map.  There is a tight correlation between the 
two continuum emissions at 24~{\micron} and 88~{\micron} for about two 
orders of magnitude. Thus the 88~{\micron} continuum signal detected with 
FIS-FTS reasonably traces dust emission.  As shown in the plot, brighter 
regions exhibit a little bluer colors, indicating hotter dust temperatures, 
and thus stronger radiation field.  All the data points in every group are 
on almost the same relation, suggesting that the properties of dust are 
similar throughout the 30 Dor region.

\section{Discussion}

We reveal the extended distribution of the [OIII] emission around the 30 
Dor region; the distribution is found to be much wider than the boundary of 
the previously observed area.  Even on the peripheral regions $4'-5'$ away 
from the center of 30 Dor, we obtain [OIII] line intensities so high as 
$(1-2)\times10^{-6}$ W m$^{-2}$ sr$^{-1}$.  From calculation of a 
three-level system for the lowest energy levels of O$^{2+}$ 
\citep{Mendoza1983}, 
we estimate that the gas densities of $60-100$ cm$^{-3}$ are required to 
explain the above line intensities.  Here we assume that 50 \% of oxygen 
is in the ionization state of O$^{2+}$ and the O/H abundance ratio is 
$2.3\times10^{-4}$ \citep{Madden2006}. For the depth of the emitting 
region, we adopt 75 pc (=$5'$), which is similar to the distance from 
the center of 30 Dor.

To ionize O$^{+}$ to O$^{2+}$, energetic photons ($>$ 35.12 eV) are 
required, which can be produced only by very massive stars such as early O 
type stars and Wolf-Rayet (WR) stars.  We estimate the contribution of the 
massive stars to the ionization of O$^{+}$.  R136 has a lot of massive 
stars in the compact central region, as revealed by the Hubble Space 
Telescope (\cite{Selman1999}; \cite{Massey1998}).  The ionization structure 
at 30 Dor (a few arc-minutes around R136) is determined by a balance between 
radiation from massive stars in R136 and the distribution of the 
interstellar medium (\cite{Poglitsch1995}; \cite{Indebetouw2009}).

\begin{figure}
  \begin{center}
    \FigureFile(80mm,85mm){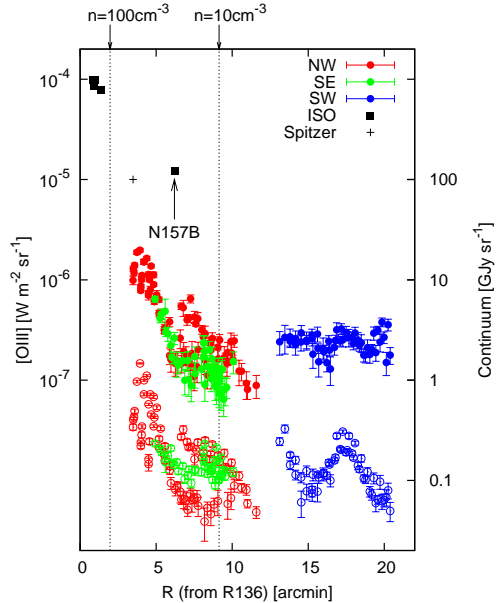}
    %%% \FigureFile(width,height){filename}
  \end{center}
  \caption{Radial profiles of the distributions of the [OIII] line emission 
  	(filled circles) and far-IR continuum (open circles) plotted against 
  	the distance from the super star cluster R136. ISO/LWS observations of 
  	30 Dor are plotted with the filled squares as a reference of the 
  	central region \citep{Vermeij2002a}. Note that the ISO/LWS data point 
  	labeled N157B is located on the supernova remnant NGC2060 (see figure 
  	\ref{fig:obs_area}).  
	The plus (+) indicates the result of the Spitzer/MIPS observation 
	in the SED mode \citep{Indebetouw2009}.
	Str\"{o}mgren radii of R136 for ambient gas 
  	densities of 100 cm$^{-3}$ and 10 cm$^{-3}$ are plotted by the 
  	dotted lines.}
  \label{fig:r_depend}
\end{figure}

To assess the influence range of radiation field from the R136 super star 
cluster, we investigate the change of the [OIII] intensity with distance 
from R136.  Figure \ref{fig:r_depend} shows the radial profiles of the 
distributions of the [OIII] emission and the far-IR continuum emission 
plotted against the distance from R136.  The upper plots (filled circles) 
show the [OIII] line emission profile, while the lower plots (open circles) 
show the far-IR continuum emission profile. The SR group is excluded in 
this plot.  The [OIII] emission and far-IR continuum of the NW and SE 
groups show a clear dependence on the distance from R136 as seen in 
figure \ref{fig:r_depend}, while those of the SW group do not; the latter 
is likely to have local ionization sources other than R136.

Figure \ref{fig:r_depend} shows the wide distribution of the highly-ionized 
region around the central super star cluster R136.  The influence of R136 
seems to reach $\sim10'$, or 150 pc away from R136.  The 
intensities of the central [OIII] emission detected with the ISO/LWS are 
consistent with the extrapolation of the AKARI/FIS-FTS data to the inner 
area of $1'$ from R136, as shown in figure \ref{fig:r_depend}.  The ISO/LWS 
data point labeled N157B is located on the supernova remnant NGC2060 (see 
figure \ref{fig:obs_area}).  It is notable that the far-IR continuum 
emission has a radial dependence similar to the [OIII] emission.  The 
vertical line in the figure indicates the Str\"{o}mgren radius of hydrogen 
gas (about $2'$) determined by the R136 super star cluster.  To calculate 
the value, we assume that the massive stars within a radius of 15 pc ($1'$) 
(\cite{Bonanos2009}; \cite{Selman1999}; \cite{Massey1998}) concentrate on 
the center and that ambient gas density and temperature are 100 cm$^{-3}$ 
and $10^4$ K, respectively. 
The gas density adopted here reflects the above value that is estimated 
from the [OIII] line intensity observed with FIS-FTS.
These gas parameters are also reasonable from the results obtained with 
the radio \citep{Peck1997}, 
ISO \citep{Vermeij2002b}, and Spitzer observations \citep{Indebetouw2009}, 
which suggested densities around 200 cm$^{-3}$ near the center of 30 Dor.  
We consider a conservative case that the gas density is constant at the value 
estimated in the peripheral region. 
Yet, from figure \ref{fig:r_depend}, it is clear that the simple calculation 
cannot explain the widely-extended distribution of the [OIII] line 
emission.  
The central bubble of 1$'$ in diameter would increase the 
Str\"{o}mgren radius, but its presence does not dramatically change the 
result.  
To explain the observed distribution, we have to assume a lot of 
hidden massive stars, and/or non-uniform distribution of the interstellar 
media, that is, clumpy distribution and leakage of ionizing photons.  It is 
not likely that a lot of embedded massive stars are hidden in the field.  
If the gas density is reduced down to 10 cm$^{-3}$, the Str\"{o}mgren 
radius expands to $9'$, which is comparable to the observed size of the 
distribution of the [OIII] emission.  Therefore a picture of the clumpy 
distribution of the interstellar media with a volume filling factor of 
$\sim 0.1$ is a plausible explanation.

\begin{figure}
  \begin{center}
    \FigureFile(120mm,150mm){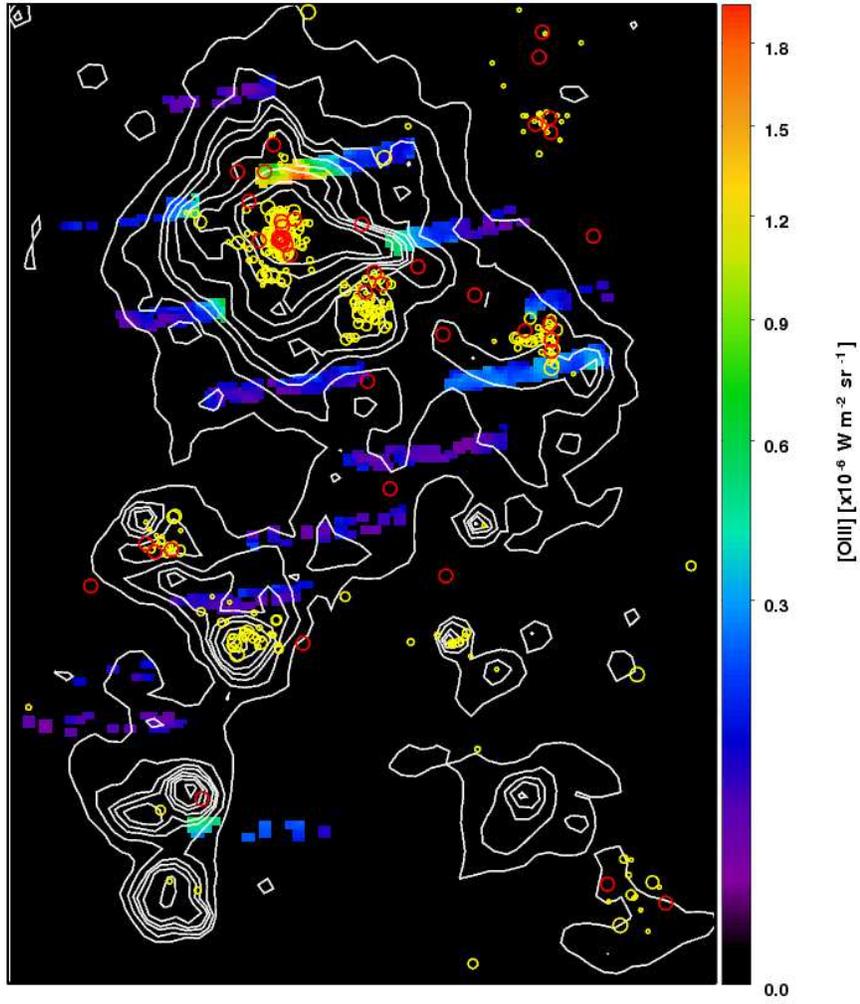}
    %%% \FigureFile(width,height){filename}
  \end{center}
  \caption{Sizes of the Str\"{o}mgren spheres for cataloged Wolf-Rayet 
  	stars (red or dark circles) and O2--O7 stars (yellow or bright 
	circles) calculated under the assumption of n = 100 cm$^{-3}$ 
	and T = 10$^{4}$ K for ambient 
  	gas. The [OIII] intensity map is the same as in figure 
  	\ref{fig:OIII_map}, but the contours show the H$\alpha$ emission 
  	\citep{Gaustad2001}.}
  \label{fig:stromgren}
\end{figure}

Thus it is difficult to explain the ionization in the areas far away from 
30 Dor solely by radiation from massive stars in R136.  However many other 
massive stars are also distributed widely in the 30 Dor region in addition 
to those in R136. \citet{Bonanos2009} presented a catalog of 1750 massive 
stars in the LMC with accurate spectral types; stars in the R136 super star 
cluster are not included in the catalog.  We calculated the Str\"{o}mgren 
radius of hydrogen gas for each cataloged star \citep{Vacca1996}, assuming 
the physical condition of $100$ cm$^{-3}$ and $10^4$ K for gas density and 
temperature, respectively. The gas density again corresponds to the value 
obtained from the observed [OIII] line intensity.  The Str\"{o}mgren radius 
thus obtained is plotted in figure \ref{fig:stromgren} with the circle for 
each massive star.  So far the Str\"{o}mgren radii are calculated for 
H$^{+}$ ions, but not for O$^{2+}$ ions.  However our simplified 
calculation shows that the Str\"{o}mgren radius for O$^{2+}$ is almost the 
same as for H$^{+}$ as long as ionizing photons have energies significantly 
higher than the O$^{+}$ ionization energy (35.12 eV).  At least, a 
Str\"{o}mgren radius for O$^{2+}$ is no larger than that for H$^{+}$ ions.  
Thus it is difficult to explain the observed [OIII] emission in the outer 
region of 30 Dor by radiation from the distributed massive stars, too.

\begin{figure}
  \begin{center}
    \FigureFile(80mm,75mm){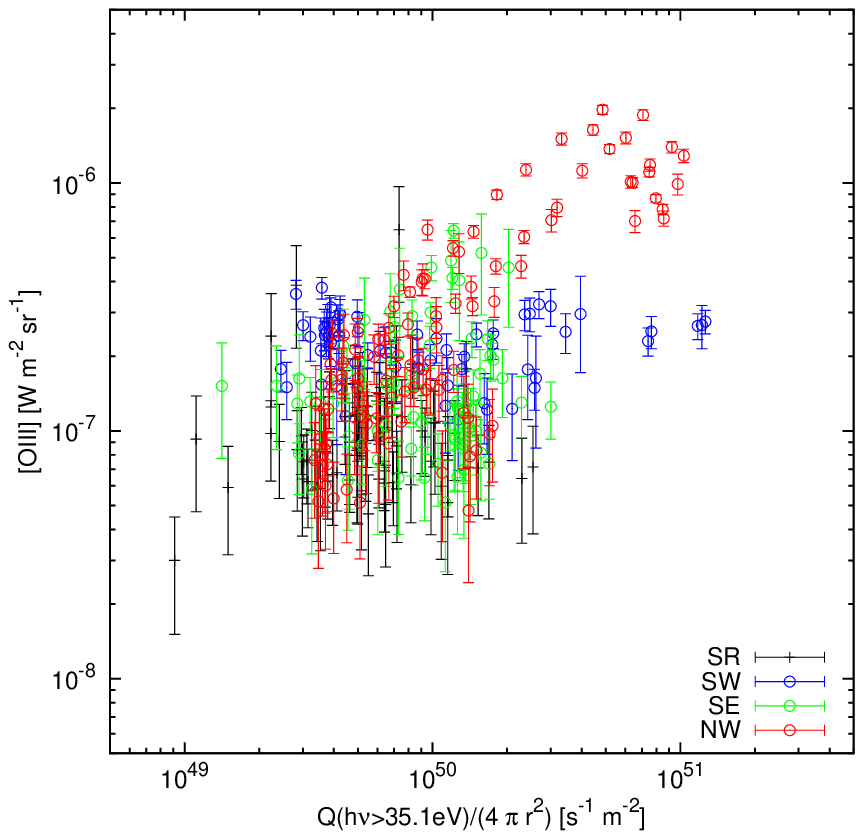}
    %%% \FigureFile(width,height){filename}
  \end{center}
  \caption{[OIII]~88{\micron} line intensity plotted as a function 
	of the O$^{+}$-ionizing ($>$ 35.12 eV) photon flux, which is 
	calculated for each observing point by summing the contribution of 
	the cataloged massive stars shown in figure \ref{fig:stromgren} 
	without correcting for any absorptions. 
	}
  \label{fig:Qi_OIII}
\end{figure}

In order to evaluate the contribution of hidden sources of ionization to 
the observed distribution of the [OIII] line emission, in 
figure \ref{fig:Qi_OIII}, we plot the [OIII] line intensity as a function 
of the O$^{+}$-ionizing photon flux expected from the cataloged stars. We 
obtain the flux at each position by adding the total absorption-free photon 
fluxes coming from the cataloged WR stars and O2--O7 stars shown in 
figure \ref{fig:stromgren}. Here we use the projected distance for each 
star.  As seen in the figure, there is a correlation between the 
[OIII] line intensity and the absorption-free ionizing flux, at least for 
the NW region and part of the SE region. This suggests 
that the ISM is very clumpy in those regions and that there are no 
significant hidden sources of ionization other than the dense star clusters 
in figure \ref{fig:stromgren}.  An exception is the SW region that contains 
30 Dor C, where the [OIII] line emission shows almost no dependence on the 
ionizing photon flux with line intensities above and below the correlation 
at smaller and larger photon fluxes, respectively. The lack of the 
correlation suggests the presence of another diffuse ionizing source, which 
can be attributed to hard radiation and/or shocks from the supernova remnants 
30 Dor C. As for the lower line intensities at photon larger fluxes, some 
areas of the SW region are so closely located near massive stars (see 
figure \ref{fig:stromgren}) that the photon fluxes calculated from the 
projected distances are very likely overestimated. 

\begin{figure}
  \begin{center}
    \FigureFile(80mm,75mm){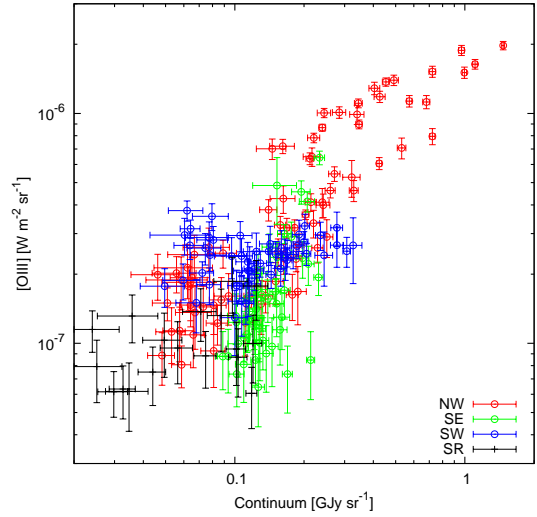}
    %%% \FigureFile(width,height){filename}
  \end{center}
  \caption{Scatter plot between the [OIII] line intensity and the far-IR 
  	continuum emission at 88~{\micron}, both of which are obtained with 
  	FIS-FTS.  The colors of data points are the same as in figure 
  	\ref{fig:corr_Ha}.}
  \label{fig:corr_cnt}
\end{figure}

Figure \ref{fig:corr_cnt} shows a scatter plot between the [OIII] line and 
the 88~{\micron} continuum emission, both obtained with FIS-FTS, where we 
see an overall correlation between them.  It is consistent with the similar 
trend of radial dependence between the [OIII] emission and the continuum 
emission as seen in figure \ref{fig:r_depend}.  This correlation is 
interesting because they are likely of different origins; the far-IR 
continuum represents the dust emission mostly coming from neutral regions, 
while the [OIII] emission originates in highly-ionized gas regions.  Thus 
the correlation suggests that the highly-ionized gas and dust are mixed 
well and exist in the same region, or have a clumpy structure where the 
dust is shielded from energetic photons by local dense gases. This 
interpretation is consistent with the widely extended distribution of the 
[OIII] emission. 

The ionization of the SW region does not seem to be related to the central 
sources in the 30 Dor region as seen in figure \ref{fig:r_depend}. In 
figure \ref{fig:corr_cnt}, all the data points seem to follow the global 
trend of correlation.  By looking locally in detail, however, the SW group 
has a much weaker dependence of the [OIII] intensity on the far-IR 
continuum emission, i.e. no clear correlation between the [OIII] and the 
dust emission, suggesting that the dust and ionized gases are not mixed 
well in this region. 
As described above, most data points of the SW region are located on the 
supershell of the supernova remnants 30 Dor C \citep{Townsley2006}. Then, 
a large fraction of dust in the supershell may have been destroyed by the 
supernovae.

\section{Summary}

We have performed spectroscopic surveys of the LMC with AKARI/FIS-FTS with 
a good spatial resolution of about $1'$.  We have analyzed spectral data 
from the 14 observational points distributed widely around the 30 Dor 
region on a degree scale.  We find that the [OIII]~88{\micron} emission is 
widely extended by more than $10'$ (150 pc) from the central super star 
cluster, R136.  Since the [OIII] line traces highly-ionized gas regions 
heated by very massive stars, the detection of widely extended [OIII] 
emission suggests that energetic photons are escaping into regions far from 
massive stars localized at the center of the 30 Dor region. It is unlikely 
that a lot of embedded massive stars are hidden in the outer region of 30 
Dor.  Therefore, a plausible interpretation is that the interstellar medium 
has a patchy or porous structure around the 30 Dor region, so that 
high-energy photons can pervade into wider areas. This is also supported by 
a good correlation between the [OIII] emission and the far-IR continuum 
emission, suggesting that the gas and dust are well mixed in the 
highly-ionized region where the dust survives in clumpy dense clouds 
shielded from the energetic photons.

\bigskip

We would express many thanks to an anonymous referee for giving us important 
comments.
We thanks all members of the AKARI project. Particularly, we express our 
gratitude to the AKARI data reduction team for their dedicated work in 
generating the TSD, and developing data analysis pipelines. AKARI is a 
JAXA project with the participation of ESA.  This work is based in part 
on observations made with the Spitzer Space Telescope, which is operated 
by the Jet Propulsion Laboratory, California Institute of Technology under 
a contract with NASA.  This work also made use of data product from the 
Southern H-Alpha Sky Survey Atlas (SHASSA), which is supported by the 
National Science Foundation. 

%%%
% See the manual for the detail.
%%%


\begin{thebibliography}{}
% Journals(e.g. A\&A,ApJ,AJ,NMRAS,PASP ...)
% Authors, Year, Journal, Vol#, Page#
% Journal Title Abbreviation >> http://www.asj.or.jp/pasj/Jabb.html

\bibitem[Bamba et al.(2004)]{Bamba2004}
  Bamba, A., Ueno, M., Nakajima, H., \& Koyama, K.\ 2004, \apj, 602, 257-263

\bibitem[Bonanos et al.(2009)]{Bonanos2009}
  Bonanos, A. Z. et al.\ 2009, \aj, 138, 1003-1021
%  Bonanos, A.Z., Massa, D.L., Sewilo, M., Lennon, D.J., Panagia, N., Smith, L.J., 
%  Meixner, M., Babler, B.L., Bracker, S., Meade, M.R., Gordon, K.D., Hora, J.L., 
%  Indebetouw, R. \& Whitney, B.A.\ 2009, \aj, 138, 1003-1021

\bibitem[Gaustad et al.(2001)]{Gaustad2001}
  Gaustad, J. E., McCullough, P. R., Rosing, W., \& Van Buren, D.\ 2001, \pasp, 113, 1326-1348

\bibitem[Indebetouw et al.(2009)]{Indebetouw2009}
  Indebetouw, R. et al.\ 2009, \apj, 694, 84-106
%  Indebetouw, R., de Messie\`res, G.E., Madden, S., Engelbracht, C., Smith, J.D., Meixner, M., 
%  Brandl, B., Smith, L.J., Boulanger, F., Galliano, F., Gordon, K., Hora, J.L., Sewilo, M., 
%  Tielens, A.G.G.M., Werner, M. \& Wolfire, M.G.\ 2009, \apj, 694, 84-106

\bibitem[Kawada et al.(2007)]{Kawada2007}	
  Kawada, M. et al.\ 2007, \pasj, 59, S389-S400
%  Kawada, M., Baba, H., Barthel, P. D., Clements, D., Cohen, M., Doi, Y., Figueredo, E., 
%  Fujiwara, M., Goto, T., Hasegawa, S., Hibi, Y., Hirao, T., Hiromoto, N., Jeong, W.-S., 
%  Kaneda, H., Kawai, T., Kawamura, A., Kester, D., Kii, T., Kobayashi, H., Kwon, S. M., 
%  Lee, H. M., Makiuti, S., Matsuo, H., Matsuura, S., Meller, T. G., Murakami, N., Nagata, H., 
%  Nakagawa, T., Narita, M., Noda, M., Oh, S. H., Okada, Y., Okuda, H., Oliver, S., Ootsubo, T., 
%  Pak, S., Park, Y.-S., Pearson, C. P., Rowan-Robinson, M., Saito, T., Salama, A., Sato, S., 
%  Savage, R. S., Serjeant, S., Shibai, H., Shirahata, M., Sohn, J., Suzuki, T., Takagi, T., 
%  Takahashi, H., Thomson, M., Usui, F., Verdugo, E., Watabe, T., White, G. J., Wang, L., 
%  Yamamura, I., Yamauchi, C. \& Yasuda, A.\ 2007, \pasj, 59, S389-S400

\bibitem[Kawada et al.(2008)]{Kawada2008}
  Kawada, M. et al.\ 2008, \pasj, 60, S389-S397
%  Kawada, M., Takahashi, H., Murakami, N., Matsuo, H., Okada, Y., Yasuda, A., 
%  Matsuura, S., Shirahata, M., Doi, Y., Kaneda, H., Ootsubo, T., Nakagawa, T., 
%  \& Shibai, H.\ 2008, \pasj, 60, S389-S397

\bibitem[Kennicutt(1984)]{Kennicutt1984}
  Kennicutt, R. C., Jr.\ 1984, \apj, 287, 116-130

\bibitem[Madden et al.(2006)]{Madden2006}
  Madden, S. C., Galliano, F., Jones, A. P., \& Sauvage, M.\ 2006, \aap, 446, 877-896

\bibitem[Massey \& Hunter(1998)]{Massey1998}
  Massey, P., \& Hunter, D. A.\ 1998, \apj, 493, 180-194

\bibitem[Meixner et al.(2006)]{Meixner2006}	
  Meixner, M. et al.\ 2006, \aj, 132, 2268-2288
%  Meixner, M., Gordon, K. D., Indebetouw, R., Hora, J.L., Whitney, B., Blum, R., Reach, W., 
%  Bernard, J.-P., Meade, M., Babler, B., Engelbracht, C.W., For, B.-Q., Misselt, K., 
%  Vijh, U., Leitherer, C., Cohen, M., Churchwell, Ed B., Boulanger, F., Frogel, J.A., 
%  Fukui, Y., Gallagher, J., Gorjian, V., Harris, J., Kelly, D., Kawamura, A., Kim, S., 
%  Latter, W.B., Madden, S., Markwick-Kemper, C., Mizuno, A., Mizuno, N., Mould, J., 
%  Nota, A., Oey, M.S., Olsen, K., Onishi, T., Paladini, R., Panagia, N., Perez-Gonzalez, P., 
%  Shibai, H., Sato, S., Smith, L., Staveley-Smith, L., Tielens, A.G.G.M., Ueta, T., 
%  van Dyk, S., Volk, K., Werner, M. \& Zaritsky, D.\ 2006, \aj, 132, 2268-2288

\bibitem[Mendoza(1983)]{Mendoza1983}
  Mendoza, C.\ 1983, in Planetary nebulae; Proceedings of the Symposium, London, 
  	England, August 9-13, 1982,
	(D. Reidel Publishing Co., Dordrecht) 143-172.

\bibitem[Murakami et al.(2007)]{Murakami2007}	
  Murakami, H. et al.\ 2007, \pasj, 59, S369-S376
%  Murakami, H., Baba, H., Barthel, P., Clements, D. L., Cohen, M., Doi, Y., Enya, K., 
%  Figueredo, E., Fujishiro, N., Fujiwara, H., Fujiwara, M., Garcia-Lario, P., Goto, T., 
%  Hasegawa, S., Hibi, Y., Hirao, T., Hiromoto, N., Hong, S. S., Imai, K., Ishigaki, M., 
%  Ishiguro, M., Ishihara, D., Ita, Y., Jeong, W.-S., Jeong, K. S., Kaneda, H., Kataza, H., 
%  Kawada, M., Kawai, T., Kawamura, A., Kessler, M. F., Kester, D., Kii, T., Kim, D. Chan, 
%  Kim, W., Kobayashi, H., Koo, B. C., Kwon, S. M., Lee, H. M., Lorente, R., Makiuti, S., 
%  Matsuhara, H., Matsumoto, T., Matsuo, H., Matsuura, S., Meller, T. G., Murakami, N., 
%  Nagata, H., Nakagawa, T., Naoi, T., Narita, M., Noda, M., Oh, S. H., Ohnishi, A., 
%  Ohyama, Y., Okada, Y., Okuda, H., Oliver, S., Onaka, T., Ootsubo, T., Oyabu, S., Pak, S., 
%  Park, Y.-S., Pearson, C. P., Rowan-Robinson, M., Saito, T., Sakon, I., Salama, A., 
%  Sato, S., Savage, R. S., Serjeant, S., Shibai, H., Shirahata, M., Sohn, J., Suzuki, T., 
%  Takagi, T., Takahashi, H., Tanabe, T., Takeuchi, T. T., Takita, S., Thomson, M., 
%  Uemizu, K., Ueno, M., Usui, F., Verdugo, E., Wada, T., Wang, L., Watabe, T., Watarai, H., 
%  White, G. J., Yamamura, I., Yamauchi, C. \& Yasuda, A.\ 2007, \pasj, 59, S369-S376

\bibitem[Murakami et al.(2010)]{Murakami2010}
  Murakami, N. et al.\ 2010, \pasj, 62, 1155-1166
%  Murakami, N., Kawada, M., Ootsubo, T., Okada, Y., Takahashi, H., Yasuda, A., Kaneda, H., 
%  Matsuo, H., Baluteau, J.-P., Davis-Imhof, P., Gom, B.G., Naylor, D.A., Zavagno, A., 
%  Yamamura, I., Matsuura, S., Shirahata, M., Doi, Y., Nakagawa, T. \& Shibai, H.\ 2010, 
%  \pasj, 62, 1155-1166

\bibitem[Onaka et al.(2007)]{Onaka2007}	
  Onaka, T. et al.\ 2007, \pasj, 59, S401-S410
%  Onaka, T., Matsuhara, H., Wada, T., Fujishiro, N., Fujiwara, H., Ishigaki, M., Ishihara, D., 
%  Ita, Y., Kataza, H., Kim, W., Matsumoto, T., Murakami, H., Ohyama, Y., Oyabu, S., Sakon, I., 
%  Tanabe, T., Takagi, T., Uemizu, K., Ueno, M., Usui, F., Watarai, H., Cohen, M., Enya, K., 
%  Ootsubo, T., Pearson, C. P., Takeyama, N., Yamamuro, T. \& Ikeda, Y.\ 2007, 
%  \pasj, 59, S401-S410

\bibitem[Peck et al.(1997)]{Peck1997}
  Peck, A. B., Goss, W. M., Dickel, H. R., Roelfsema, P. R., Kesteven, M. J., 
  Dickel, J. R., Milne, D. K., \& Points, S. D.\ 1997, \apj, 486, 329-337

\bibitem[Poglitsch et al.(1995)]{Poglitsch1995}
  Poglitsch, A., Krabbe, A., Madden, S. C., Nikola, T., Geis, N., 
  Johansson, L. E. B., Stacey, G. J., \& Sternberg, A.\ 1995, 
  \apj, 454, 293-306

\bibitem[Rubin et al.(2009)]{Rubin2009}
  Rubin, D. et al.\ 2009, \aap, 494, 647-661
%  Rubin, D., Hony, S., Madden, S.C., Tirlrns, A.G.G.M., Meixner, M., Indebetouw, R., Reach, W., 
%  Ginsburg, A., Kim, S., Mochizuki, K., Babler, B., Block, M., Bracker, S.B., Engelbracht, C.W., 
%  For, B.-Q., Gordon, K., Hora, J.L., Leitherer, C., Meade, M., Misselt, K., Sewilo, M., 
%  Vijh, U., Whitney, B.\ 2009, \aap, 494, 647-661

\bibitem[Rubio et al.(1998)]{Rubio1998}
  Rubio, M., Barb\'{a}, R. H., Walborn, N. R., Probst, R. G., Garc\'{i}a, J., 
  \& Roth, M. R.\ 1998, \aj, 116, 1708-1718

\bibitem[Selman et al.(1999)]{Selman1999}
  Selman, F., Melnick, J., Bosch, G., \& Terlevich, R.\ 1999, 
  \aap, 347, 532-549

\bibitem[Smith et al.(2000)]{Smith2000}
  Smith, R. C., Leiton, R., \& Pizarro, S.\ 2000, 
  in Stars, gas and dust in galaxies: Exploring the Links, 
  ASP Conference Proceedings, Vol. 221. Ed. Alloin, D., Olsen, K., 
  and Galaz, G., San Francisco, 83-86.

\bibitem[Townsley et al.(2006)]{Townsley2006}
  Townsley, L. K., Broos, P. S., Feigelson, E. D., Brandl, B. R., Chu, Y.-H., 
  Garmire, G. P., \& Pavlov, G. G.\ 2006, \aj, 131, 2140-2163

\bibitem[Vacca et al.(1996)]{Vacca1996}
  Vacca, W. D., Garmany, C. D., \& Shull, J. M.\ 1996, \apj, 460, 914-931

\bibitem[Vermeij et al.(2002)]{Vermeij2002a}
  Vermeij, R., Damour, F., van der Hulst, J. M., \& Baluteau, J-.P.\ 2002, 
  \aap, 390, 649-665

\bibitem[Vermeij \& van der Hulst(2002)]{Vermeij2002b}
  Vermeij, R., \& van der Hulst, J. M.\ 2002, \aap, 391, 1081-1095

\bibitem[Yamaguchi et al.(2009)]{Yamaguchi2009}
  Yamaguchi, H., Bamba, A., \& Koyama, K.\ 2009, \pasj, 61, S175-S181

%\bibitem[Aauthor \& Bauthor(2001b)]{key-3}
%  Aauthor, A., Bauthor, B.\ 2001, Name of Book(Publisher, Tokyo) page0
%......
% Editorial Books
%\bibitem[Dauthor(2001)]{key-n}
%  Dauthor A.~A.\ 2001, in Name of Book,
%   ed.\  D.~Editor (Publisher, Tokyo) page0
%
\end{thebibliography}
\end{document}